\documentstyle[aps,prl,graphicx]{revtex}

\input psfig.sty
\draft
\begin{document}
\twocolumn 
\wideabs{  
\title{Observation of Tkachenko Oscillations in Rapidly Rotating Bose-Einstein Condensates}
\author{I. Coddington, P. Engels, V. Schweikhard, and E.~A. Cornell\cite{qpdNIST}}
\address{JILA, National Institute of Standards and Technology and University of Colorado,
and Department of Physics, University of Colorado, Boulder,
Colorado 80309-0440}
\date{\today}

\maketitle

\begin{abstract}

We directly image Tkachenko waves in a vortex lattice in a
dilute-gas Bose-Einstein condensate.  The low (sub-Hz) resonant
frequencies are a consequence of the small but nonvanishing
elastic shear modulus of the vortex-filled superfluid. The
frequencies are measured for rotation rates as high as 98\% of
the centrifugal limit for the harmonically confined gas.
Agreement with a hydrodynamic theory worsens with increasing
rotation rate, perhaps due to the increasing fraction of the
volume displaced by the vortex cores. We also observe two
low-lying m=0 longitudinal modes at about 20 times higher
frequency.
\end{abstract}

\pacs{03.75.Lm,67.90.+z,67.40.Vs,32.80.Pj}
} 
\par

\par
We have all seen a cylindrically confined fluid support
azimuthal flow whether we are watching water flow down a drain
or a recently stirred cup of coffee.  What is somewhat harder to
imagine is a fluid sustaining oscillatory azimuthal flow.
Instinctively one does not expect a fluid to support shear
forces, and this would seem especially true in the case of
zero-viscosity superfluids, but such intuition is incomplete.
\par
The key issue is vortices. In 1955, Feynman
\cite{Feynman} predicted that a superfluid can rotate when
pierced by an array of quantized singularities or vortices. In
1957, Abrikosov \cite{Abrikosov} demonstrated that such vortices
in a type II superconductor will organize into a triangular
crystalline lattice due to their mutual repulsion.  Not
surprisingly, the Abrikosov lattice has an associated rigidity.
In 1966, Tkachenko proposed that a vortex lattice in a
superfluid would support transverse elastic modes
\cite{Tkachenko}.  First observed by Andereck et al.\cite{Andereck}, Tkachenko oscillations have
been the object of considerable experimental and theoretical
effort in superfluid helium, much of which was summarized by
Sonin in 1987 \cite{Sonin}.


\par
In the last two years it has become possible to achieve a vortex
lattice state in dilute gas BEC
\cite{Dalibard,Ketterle,Foot,Jilanucleation} and recent
theoretical work \cite{Anglin} has suggested that Tkachenko
oscillations are also attainable. In this Letter we report the
observation of Tkachenko oscillations in BEC.  The particular
strengths of BEC are that in the clean environment of a
magnetically trapped gas there is no vortex pinning, and
spatiotemporal evolution of the oscillation may be directly
observed.  Since the original submission of this paper Gordon
Baym and Baksmaty et al. have independantly published
theoretical works\cite{Baym,Bigelow} that precisely describe our
data.

\par
We begin the experiment with a rotating condensate held in an
axially symmetric trap with trap frequencies
$\{\omega_{\rho},\omega_{z}\}=2\pi\{8.3,5.2\}$~Hz.  The
condensed cloud contains 1.5-2.9~million $^{87}Rb$ atoms in the
$|F=1,m_{F}=-1\rangle$ state.  The cloud rotates about the
vertical, z~axis. Rotation rates for the experiments described
in this paper range from $\Omega$=0.84 to $\Omega$=0.975
($\Omega$ defined as condensate rotation rate over
${\omega_{\rho}}$). We have no observable normal cloud implying
a T/Tc $< 0.6$. The means by which we prepare this condensate is
identical to our previous work
\cite{Jilanucleation,Jilastripes}.  As before, rotation can be
accurately measured by comparing the condensate aspect ratio to
the trap aspect ratio. Vortices, which are too small to observe
in trap, can be seen by turning off the trap and allowing the
cloud to expand to five times its original size, or typically
$380$~$\mu$m FWHM, and imaging along the direction of
rotation\cite{GiantVortex}.  At our high rotation rates the
condensate is oblate and the vortex cores are essentially
vertical lines except right at the surface.

\par
We excite lattice oscillations by two mechanisms. The first
mechanism presented is based on the selective removal of atoms
that has also been discussed in previous work
\cite{GiantVortex}.  With this method we remove
atoms at the center of the condensate with a resonant, focused
laser beam sent through the condensate along the axis of
rotation. The width of the ``blasting" laser beam is $16$~
$\mu$m FWHM (small compared to an in-trap condensate FWHM of
$75$~$\mu$m), with a Gaussian intensity profile. The frequency
of the laser is tuned to the $F''=1\rightarrow F' = 0$
transition of the D2 line, and the recoil from a spontaneously
scattered photon blasts atoms out of the condensate.  The laser
power is about 10~fW and is left on for approximately one
lattice rotation period (125~ms).

\par
\begin{figure}
\begin{center}
\psfig{figure=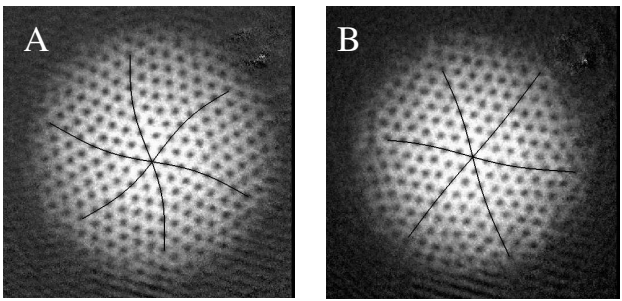,width=1\linewidth,clip=}
\end{center}
\caption {(1,0) Tkachenko mode excited by atom removal (a) taken 500~ms
after the end of the blasting pulse (b) taken 1650~ms after the
end of the blasting pulse.  BEC rotation is counterclockwise.
Lines are sine fits to the vortex lattice.}
\label{leftandrigh}
\end{figure}

The effect of this blasting laser is to remove a small (barely
observable) fraction of atoms from the center of the condensate.
This has two consequences.  First, the average angular momentum
per particle is increased by the selective removal of low
angular momentum atoms from the condensate center. This increase
then requires a corresponding increase in the equilibrium
condensate radius\cite{GiantVortex}. Secondly, the atom removal
creates a density dip in the center of the cloud. Thus, after
the blasting pulse, the condensate has fluid flowing inward to
fill the density dip and fluid flowing outward to expand the
radius. The Coriolis force acting on these flows causes the
inward motion to be diverted in the lattice rotation direction
and the outward flow to be diverted in the opposite direction.
This sheared fluid flow drags the vortices from their
equilibrium configuration and sets the initial conditions for
the lattice oscillation as can be seen from the expanded images
in Fig.~\ref{leftandrigh}.
\par
The second method of exciting the Tkachenko oscillation is
essentially the inverse of the previous method.  Instead of
removing atoms from the cloud we use a red-detuned optical
dipole potential to draw atoms into the middle of the
condensate.  To do this we focus a 850~nm laser beam onto the
condensate. The beam has $3$~$\mu$W of power and a $40$~$\mu$m
FWHM. It propagates along the direction of condensate rotation
and its effect is to create a 0.4~nK deep Gaussian dip in the
radial trapping potential. This beam is left on for 125~ms to
create an inward fluid flow similar to before. The resulting
Tkachenko oscillation was studied for $\Omega = 0.95$, and found
to be completely consistent with the atom removal method. It is
not surprising that these two methods are equivalent since one
works by creating a dip in the interaction potential and the
other creates a similar dip in the trapping potential.

\par
For these experiments, data is extracted by destructively
imaging the vortex lattice in expansion and fitting the lattice
oscillation.  To perform this fit we find a curvilinear row of
vortices going through the center of the cloud and fit a
sinewave to the locations of the vortex centers, recording the
sine amplitude. This is done for all three directions of lattice
symmetry [see Fig.~\ref{leftandrigh}], with the amplitudes
averaged to yield the net fit amplitude of the distortion.

\par
\begin{figure}
\begin{center}
\psfig{figure=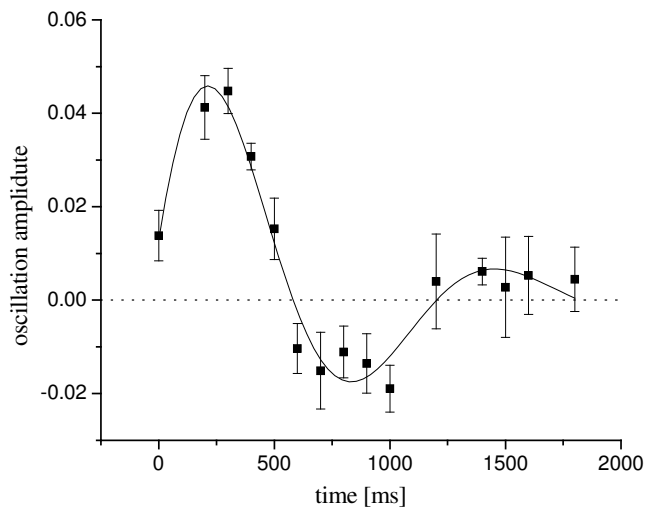,width=1\linewidth,clip=}
\end{center}
\caption {Measured oscillation amplitude for a typical excitation
$\Omega$=0.92 and $2.2\times10^{6}$ atoms. Fit is to a sinewave
times an exponential decay and yields a frequency of 0.85~Hz and
a Q of 3.  The oscillation amplitude is expressed as the average
amplitude of the sinewave fits to the vortex oscillation in
units of the radial Thomas-Fermi radius (roughly the azimuthal
displacement of a vortex a distance $0.33$~$R_{\rho}$ from the
condensate center). Both values are in expansion.}
\label{SingleOsc}
\end{figure}

The resulting oscillation [see Fig.~\ref{SingleOsc}] is heavily
damped and has a Q value of 3-5 for the data presented. Here Q
is given by Q=$2\pi f\tau_{damping}$, where $\tau_{damping}$ is
the exponential-damping time constant for the oscillation. We
are able to increase this to a Q of 10 by exciting lower
amplitude oscillations (40\% of the previous amplitude) and by
better mode matching of the blasting beam to the shape and
period of the oscillation ($40$~$\mu$m FWHM beam width and
500~ms blasting time). Measured frequencies for the
high-amplitude oscillations are the same as for the
low-amplitude, high-Q case so we do not believe that we are
seeing anharmonic shifts\cite{damping}.

\par
\begin{figure}
\begin{center}
\psfig{figure=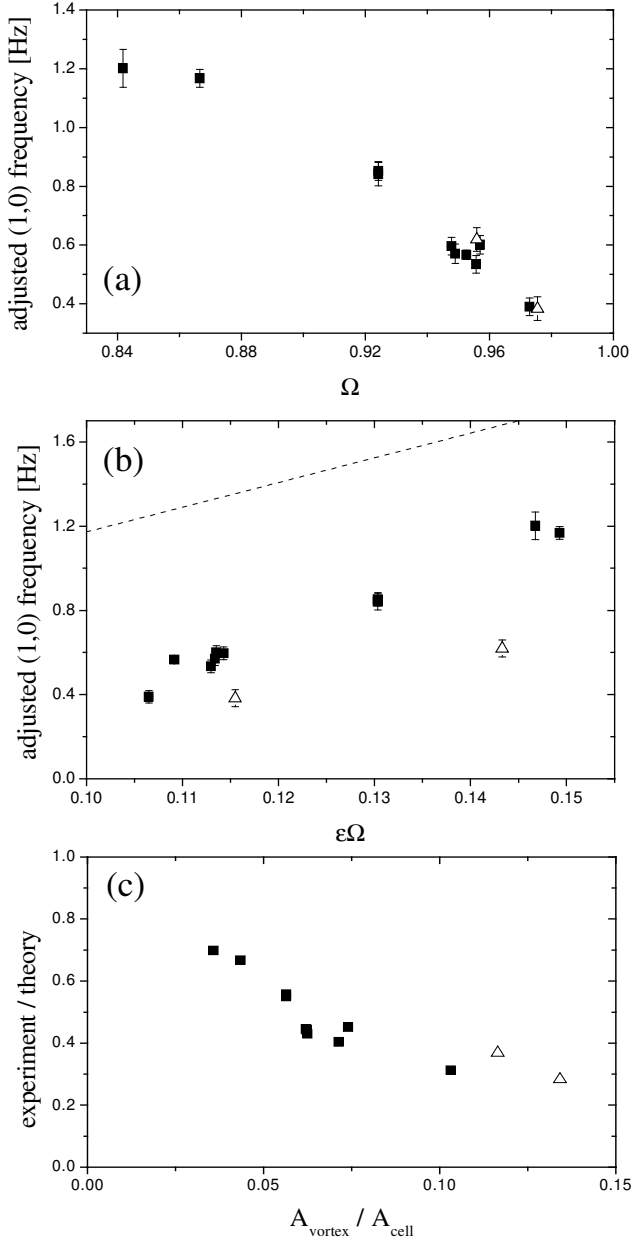,width=1\linewidth,clip=} 
\end{center}
\caption {Plot (a) shows the adjusted[15] (n=1,m=0) Tkachenko oscillation frequencies as a
function of scaled rotation rate $\Omega$. Plot (b) shows the
(1,0) frequency as a function of the theory parameter
$\epsilon\Omega$. The dotted line is the theory line
$\nu_{10}=1.43\epsilon\Omega (\frac{\omega_{\rho}}{2\pi})$ from
Ref.[10]. Note that the low number data shows much worse
agreement with theory. Plot (c) demonstrates the divergence from
theory as the ratio of vortex core area to unit cell area
increases. $A_{vortex}$ is $\pi\xi^{2}$ where the healing length
$\xi=(8\pi n a)^{-1/2}$ (here $n$ is density-weighted average
density and $a$ is the s-wave scattering length). Lattice cell
area $A_{cell}$ is $\sqrt{3}b^{2}/2$ (here b is the
nearest-neighbor vortex spacing). For all plots black squares
and triangles refer to high and low atom number experiments,
respectively. }
\label{CompAnglin}
\end{figure}

Because of the characteristic s-bend shape and the low resonant
frequency of these oscillations [see Fig.~\ref{CompAnglin}(a)]
we interpret them to be the (n=1,m=0) Tkachenko oscillations
predicted by Anglin\cite{Anglin}. Here (n,m) refer to the radial
and angular nodes, respectively, in the presumed quasi-2-D
geometry. The calculations of Ref.\cite{Anglin} predict that
these lattice oscillations should have a frequency of
$\nu_{10}=1.43\epsilon\Omega (\frac{\omega_{\rho}}{2\pi})$ for
the (1,0) mode and $\nu_{20}=2.32\epsilon\Omega
(\frac{\omega_{\rho}}{2\pi})$ for the (2,0) mode.  Here
$\epsilon=b/R_{\rho}$ denotes the nearest-neighbor vortex
spacing, b, over the radial Thomas-Fermi radius, $R_{\rho}$. For
our system these predicted frequencies are around 1-2~Hz and are
therefore far slower than any of the density-changing coherent
oscillations of the condensate except for the m=-2 surface
wave\cite{Jilanucleation,Dalibard2,Sandro,Fetter}. In addition
the shape of the observed oscillation agrees well with theory.
Specifically, the prediction\cite{Anglin} that the spatial
period of a sinewave fit to a row of vortices in a (1,0)
oscillation should be 1.33~$R_{\rho}$ is in perfect agreement
with our data.

\par
The predicted frequencies are however problematic.  To make the
comparison to the theory presented in Ref.\cite{Anglin} we
excite lattice oscillations in the condensate for
$\epsilon\Omega$ ranging from 0.10 to 0.15. This is achieved by
varying number and rotation rate. Over this range of
$\epsilon\Omega$ the oscillation frequencies measured are
consistently lower than those predicted by theory as can be seen
in Fig.~\ref{CompAnglin}(b). For the slowest rotations,
$\Omega$=0.84 ($\epsilon\Omega=0.15$,N=$2.5\times10^{6}$), we
observe frequencies that are as close as 0.70 of the predicted
value. However, at larger rotation rates, $\Omega$=0.975
($\epsilon\Omega=0.10$,N=$1.7\times10^{6}$), the agreement is
considerably worse (the measured value is 0.31 of the of the
predicted value). One possible explanation for this general
discrepancy is that the calculations are done in 2-D and ignore
the issues of vortex bending at the boundary and finite
condensate thickness
\cite{Anglin2}.  In those cases, however, one would expect better
agreement at high rotation rates where the condensate aspect
ratio is more 2-D. A more likely explanation is that the
continuum theory, used in the Anglin calculation, is breaking
down as the vortex core size to vortex spacing becomes finite
\cite{Anglin2}.  This suggests that at high rotation and lower
atom number we are entering a new regime.  To further explore
this possibility we reduced the atom number to
N=$7-9\times10^{5}$, while keeping $\epsilon\Omega$ roughly the
same.  This should increase the core size and exacerbate the
problem. As can be seen in Fig.~\ref{CompAnglin}(b) and
Fig.~\ref{CompAnglin}(c) the agreement with theory is
significantly worse under these conditions.

\par
\begin{figure}
\begin{center}
\psfig{figure=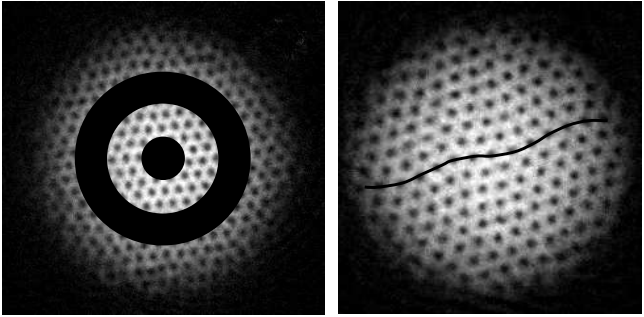,width=1\linewidth,clip=}
\end{center}
\caption {On the left are the locations where atoms are removed
from the cloud.  For the (1,0) excitations the atoms are removed
from the shaded region in the center.  For the (2,0) mode atoms
are removed from the shaded ring half a condensate radius out.
Image on the right is the resulting (2,0) mode, where the black
line has been added to guide the eye.}
\label{n2twave}
\end{figure}

We are also able to excite the (2,0) mode. We note that atom
removal creates an s-bend in the lattice that is centered on the
atom removal spot. To write two s-bends onto the lattice one
could imagine removing atoms from an annular ring instead of a
spot. To make this ring we offset the blasting beam half a
condensate radius and leave it on for 375~ms (three full
condensate rotation periods). As one can see this does lead to
an excitation of the (2,0) oscillation (see Fig.~\ref{n2twave}).
We measure the frequency of this mode as before. For 2.3 million
atoms and $\Omega = 0.95$ we measure a lattice oscillation
frequency of $1.1\pm0.1$~Hz, distinctly lower than the
theoretical prediction \cite{Anglin} of 2.2~Hz for our
parameters. It is interesting that the predicted {\it ratio} of
frequencies, $\nu_{20}/\nu_{10}$, is 1.63, in agreement with the
experimental value, $1.8\pm0.2$, measured at $\epsilon\Omega =
0.12$.

\par
Vortex motion and condensate fluid motion are intimately linked
\cite{Sonin}. In Tkachenko oscillations, the moving of vortices must
also entail some motion of the underlying fluid, and
pressure-velocity waves in the fluid must conversely entrain the
vortices. Very generally, for a substance composed of two
interpenetrating materials, one of which has an elastic shear
modulus and one of which does not (in our case, the vortex
lattice and its surrounding superfluid, respectively), one
expects to find three distinct families of sound waves in the
bulk: (i) a shear, or transverse, wave, (ii) a common-mode
pressure or longitudinal wave, and (iii) a differential
longitudinal wave, with the lattice and its fluid moving against
one another\cite{Toner}. The presence of strong Coriolis forces
makes the distinction between longitudinal and transverse waves
problematic, but the general characteristics of the three
families should extend into the rotating case. For instance, one
can still readily identify the Tkachenko modes discussed thus
far as the transverse wave. Our assumption is that the
common-mode longitudinal waves are nothing other than the
conventional hydrodynamic shape oscillations studied previously
\cite{Sandro,Fetter}.

To excite the common-mode longitudinal wave, we use the dipole
force from the 850~nm red-detuned laser described earlier. In
order to excite a broad spectrum of modes we shorten the laser
pulse to 5~ms, widen the excitation beam to a 75~$\mu$m FWHM
Gaussian profile, and increase the laser power to 1~mW,
resulting in a 30~nK deep optical potential.  We find that this
pulse excites three distinct m=0 modes: the first is the (1,0)
Tkachenko s-bend mode at about 0.6~Hz already discussed. The
second is a radial breathing mode in which the condensate radius
oscillates at $16.6\pm0.3$~Hz (or $2.0\pm0.1
\frac{\omega_{\rho}}{2\pi})$. This mode has been previously
observed\cite{Parisians}, and our observed frequency is
consistent with hydrodynamic theory for a cloud rotating at
$\Omega = 0.95$
\cite{Sandro}. As the radius of the fluid density oscillates, so
does the mean lattice spacing of the vortex lattice, but we
observe no s-type bending of the lattice at this frequency. The
fact that the frequency of the lowest m=0 radial longitudinal
mode is more than 20 times that of the transverse mode
demonstrates how relatively weak the transverse shear modulus
is.

The same laser pulse excites a third mode, at the quite distinct
frequency of $18.5\pm0.3$~Hz. This mode manifests as a rapid
s-bend distortion of the lattice indistinguishable in shape from
the 0.6~Hz (1,0) Tkachenko oscillation. 18.5~Hz is much too fast
to have anything to do with the shear modulus of the lattice,
and we were very tempted to identify this mode as a member of
the third family of sound-waves, the differential longitudinal
waves. Simulations by Cozzini and Stringari \cite{Cozzini},
however, show that our observed frequency is consistent with a
higher-order, hydrodynamic mode of the rotating fluid that can
be excited by an anharmonic radial potential such as our
Gaussian optical potential. Moreover, they show that the radial
velocity field of their mode is distorted by Coriolis forces so
as to drag the lattice sites into an azimuthally oscillating
s-bend distortion that coincidentally resembles the Tkachenko
mode. It is worth noting that without the presence of the
lattice to serve as tracers for the fluid velocity field, it
would be very difficult to observe this higher-order mode, since
this mode has very little effect on the mean radius of the
fluid. In any case, the mode at 18.5~Hz appears to be yet
another member in the family of common-mode longitudinal waves.
So far we have been unable to observe a mode we can assign to
the family of differential longitudinal waves.

\par
We would like to acknowledge James Anglin, Marco Cozzini, Sandro
Stringari John Toner and Gordon Baym for their useful
discussion. We are also appreciative of additional calculations
done by Anglin and Cozzini. The work presented in this paper was
funded by NSF and NIST.


%




\begin{thebibliography}{10}
\bibitem{damping}
Note that for all cases the reported frequency is adjusted for
damping according to the equation $f_{o}=1/2\pi ( (2\pi
f_{measured})^{2}+(1/\tau_{damping})^{2})^{1/2}$.
\end{thebibliography}
\end{document}